\documentclass{Interspeech2024}

\interspeechcameraready

\newcommand{\libtestclean}[1]{\mbox{\texttt{test-clean}}}
\newcommand{\libtestother}[1]{\mbox{\texttt{test-other}}}
\newcommand{\libtrainclean}[1]{\mbox{\texttt{train-clean-100}}}
\newcommand{\libtraincleanother}[1]{\mbox{\texttt{train-clean-360}}}
\newcommand{\africanaccented}[1]{\mbox{\texttt{African-accented English speech}}}

\title{Reduce, Reuse, Recycle: Is Perturbed Data Better than Other Language Augmentation for Low Resource Self-Supervised Speech Models}

\name{Asad}{Ullah}
\name{Alessandro}{Ragano}
\name{Andrew}{Hines}

\address{
  School of Computer Science, University College Dublin, Ireland
}
\email{asad.ullah@ucdconnect.ie, alessandro.ragano@ucd.ie, andrew.hines@ucd.ie}

\begin{document}

\maketitle

\begin{abstract}

    Self-supervised representation learning (SSRL) has demonstrated superior performance than supervised models for tasks including phoneme recognition. Training SSRL models poses a challenge for low-resource languages where sufficient pre-training data may not be available. A common approach is cross-lingual pre-training. Instead, we propose to use audio augmentation techniques, namely: pitch variation, noise addition, accented target language and other language speech to pre-train SSRL models in a low resource condition and evaluate phoneme recognition. Our comparisons found that a combined synthetic augmentations (noise/pitch) strategy outperformed accent and language knowledge transfer. Furthermore, we examined the scaling factor of augmented data to achieve equivalent performance to model pre-trained with target domain speech. Our findings suggest that for resource-constrained languages, combined augmentations can be a viable option than other augmentations.

\end{abstract}

%\section{Introduction}
\section{Introduction}
\label{sec:intro}
There are 23 out of 7000 languages spoken by more than half of the world's population~\cite{hammarstrom2005review}. Almost 40\% of languages are in danger of extinction \cite{moseley2010atlas}. Some languages such as Hokkien~\cite{ting2010language} do not have written forms.  Other languages such as Pashto~\cite{rahman1995pashto} do not have a standardized form and vary from region to region. 
%A few languages are spoken by a small population. 
These languages have limited resources regarding data, documentation, and standardized forms. Collecting standard speech datasets for these low-resource languages is challenging and still an open issue in speech processing tasks such as phoneme recognition and automatic speech recognition. 

Given these challenges, supervised learning methods typically underperform in phoneme and speech recognition in low-resource scenarios due to their reliance on large labelled datasets. In contrast, a more effective strategy often employed is self-supervised representation learning (SSRL)~\cite{Chung2019AnUA,oord2018representation,10.1109/TASLP.2021.3122291}. SSRL excels in generating general-purpose feature representations using large, unlabeled datasets~\cite{Yang2021SUPERBSP}. It achieves this by training models to learn hierarchical data representations through auxiliary tasks where the labels are generated from the data itself. For specific tasks like phoneme and speech recognition, these SSRL models are finetuned using smaller annotated datasets.

In the pre-training phase of an SSRL model, the selection of the auxiliary task and the data utilized for this task play a crucial role in determining the model's performance in downstream tasks. A common approach to tackling low-resource speech processing challenges involves developing a single, unified multilingual model rather than creating separate models for each language. This strategy is based on the premise that all languages possess a shared set of phonemes and that a larger pool of pre-training data can enhance performance in low-resource scenarios~\cite{babu2021xls,sellam2023squid}. For instance, expanding the wav2vec 2.0 architecture~\cite{baevski2020wav2vec} to 128 languages, as seen in the XLS-R model~\cite{babu2021xls}, has shown improved speech recognition performance compared to its use with English alone. However, the effectiveness of transferring phoneme sets across languages has not been thoroughly investigated. The XLS-R model, for example, includes languages like Irish, Oriya, and Manx, which only contribute a minor portion of the pre-training data. Consequently, the model's performance heavily relies on the substantial representation of dominant languages such as English or Spanish.

Deep learning models often struggle with the discrepancies between their training and testing data, known as domain mismatch. Although transferring from high-resource languages has been shown to be beneficial in low-resource scenarios, we hypothesise that it still suffers from domain mismatch. To explore this, we examine adopting data augmentation strategies as an alternative in the pre-training phase of SSRL models for low-resource scenarios. Rather than incorporating additional languages, we propose to apply data augmentation to the low-resource data in the pre-training stage. This method not only helps increase the presence of low-resource data but it also avoids real augmentation from cross-lingual pre-training data. Artificial data augmentation approaches have been extensively used to mitigate the scarcity of training data in speech processing tasks such as phoneme recognition and speech recognition. Previously data-augmentation has been used for supervised speech recognition to improve the model robustness \cite{ko2015audio}, \cite{Park2019SpecAugmentAS} and noise robustness \cite{ko2017study}. Recently, different augmentation strategies have been applied in the pre-training stage of SSRL models to improve either model robustness \cite{kharitonov2021data} or noise-robustness \cite{zhu2022noise}, \cite{wang2022improving}, \cite{huang2022spiral}. Cross-lingual augmentation strategy has been used for model robustness \cite{baevski2021unsupervised}. Recently, speed perturbation \cite{ullah2023improving} and a mix of data-augmentation \cite{Sriram2022Wav2VecAugIS} approaches have been used for SSL model robustness targeting low resource languages. However, to the best of our knowledge, a systematic analysis of the data-augmentation in pre-training stages for low resource languages compared to cross-lingual pre-training has not been undertaken. Table~\ref{tab:literature_table} presents a comparison of existing augmentation approaches and their evaluation strategies. 

\begin{table*}[!ht]
\caption{\label{tab:literature_table}Data augmentation approaches and evaluation strategies}
\centering
\begin{tabular}{p{0.6cm}|p{1.5cm}|p{0.5cm} p{0.5cm} p{0.5cm} p{0.5cm} p{0.9cm}p{1.1cm}p{1.1cm}|p{1.1cm} p{1.1cm} p{1.3cm}} %\\\hline
 % \multicolumn{9}{c}{\textbf{Approaches}}                   & \multicolumn{3}{|c}{\textbf{Evaluation Strategy}} \\\hline
 % & & & \multicolumn{6}{c}{Synth augmentation}         & &  & \\\hline
\multicolumn{9}{c}{\textbf{}}                   & \multicolumn{3}{c}{\textbf{Evaluation Strategy}} \\
Paper & Model/Arch & Pitch & Rev. & Speed & Noise & Spectral Mask & Mixed synth. aug. & Natural aug. & Model robust. & Noise robust. & Low \newline resource \\\hline
\cite{ko2015audio}             &  & & & \checkmark & & & & & \checkmark & & \\
\cite{Park2019SpecAugmentAS}   & Supervised & & & & & \checkmark &  & & \checkmark & & \\
\cite{ko2017study}             & & & & & \checkmark & & & & & \checkmark & \\\hline
\cite{kharitonov2021data}      &   & & \checkmark & \checkmark & \checkmark &  & & & \checkmark & & \\
\cite{zhu2022noise}            &   & & & & \checkmark & &  & & & \checkmark & \\
\cite{wang2022improving}       &   & & & & \checkmark & &  & & & \checkmark & \\
\cite{huang2022spiral}         & SSL  & & & & \checkmark & &  & & & \checkmark & \\
\cite{baevski2021unsupervised} &   & & & & & & & \checkmark & \checkmark & & \checkmark \\
\cite{ullah2023improving}      &   & & & \checkmark & & &  & & \checkmark & &  \\
\cite{Sriram2022Wav2VecAugIS}  &   & \checkmark & \checkmark & & \checkmark & & \checkmark & & \checkmark & & \checkmark \\\hline
\textbf{Ours}                  & SSL  & \checkmark & & & \checkmark &  & \checkmark & \checkmark  & \checkmark & & \checkmark \\\hline
\end{tabular}
\end{table*}

In this work, we explore the robustness of data-augmentation in pre-training, evaluate the performance gain due to data-augmentation in a downstream phoneme recognition task and compare the results with cross-lingual augmentation. Furthermore, we evaluate how much additional augmentation data is needed to achieve the performance equivalent to the model pre-trained with target domain speech data. We find that the performance gains for the different augmentation approaches tested were similar but that using a mixture of more than one augmentation strategy yielded better results. Furthermore, we found that the rate of improvement in classification accuracy tails off significantly from beyond 2-times augmentation.

\section{Proposed Method}
Our goal is to understand whether artificial data augmentation can be a useful approach for low-resource scenarios and compare to cross-lingual pre-training.

\noindent
\textbf{Approach} 
Our approach considers as a baseline an SSRL model with a fixed amount of pre-training data that resembles a low-resource scenario, e.g., 25 hours, and finetuning it with a further smaller labelled dataset for a downstream task. Starting from the 25 hours baseline, we do experiments by incrementally adding more data for pre-training using two different strategies: 1) Adding different languages as done in cross-lingual models such as XLS-R, and 2) Using artificial data augmentation techniques on the same 25 hours of the baseline. 
The core objective of this approach is not only to compare these two strategies but also to uncover the specific impact of adding hours of different languages on the final results, an aspect that has not been systematically explored so far. By employing this strategy, we ensure that the only variables influencing performance are the augmented data sources, whether they be from different languages or through artificial data augmentation. It’s important to note that our aim is not to identify the most optimal model or the best augmentation technique. Instead, our focus is on comparing these two methodologies and analyzing their relative effectiveness in enhancing model performance.

\noindent
 \textbf{Model:} We have chosen Autoregressive Predictive Coding (APC)~\cite{Chung2019AnUA,Chung2019GenerativePF,Chung2020VectorQuantizedAP,chung-glass-2020-improved,DBLP:journals/jstsp/YangYCGT22} as our model due to its faster training time and fewer parameters compared to other SSRL models like wav2vec 2.0~\cite{baevski2020wav2vec} or HuBERT~\cite{hsu2021hubert}. Since our focus is not on benchmarking models or identifying which model responds best to data augmentation, the specific choice of the SSRL model is of secondary importance. In general, our augmentation approach can be adapted to any SSRL model.

\noindent
\textbf{Low-Resource Language:}
We simulate our experiment using the American English language as a proxy for low-resource languages in both the pre-training and downstream phases. By doing so, we have the flexibility to compare the two strategies (artificial data augmentation and cross-lingual) with an oracle model which is obtained when using more data of the target low-resource language in the pre-training stage. We can also arbitrarily control the amount of hours we want to use in the pre-training stage.

\noindent
\textbf{Data Augmentation Strategies:}
For data augmentation in the pre-training phase, we have chosen pitch shifting and background noise addition. These techniques were selected based on previous studies indicating their effectiveness in enhancing SSRL model robustness in pre-training stages, specifically for contrastive predictive coding~\cite{kharitonov2021data} and wav2vec 2.0~\cite{Sriram2022Wav2VecAugIS}.

\begin{figure}[htb]	
	\begin{minipage}[b]{1.0\linewidth}
		\centering
		\centerline{\includegraphics[width=8.0cm]{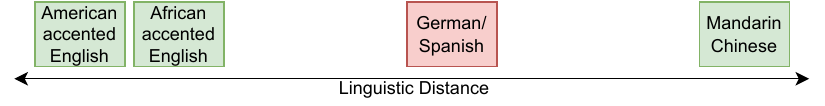}}
	\end{minipage}
 		\vspace{-6mm}
	\caption{Linguistic Distance from American English}
	\label{fig:languages}
\end{figure}

\noindent
\textbf{Cross-Lingual Pre-Training:}
To contrast data augmentation in the pre-training stage with the addition of other languages, we do not need to assess a vast array of languages. Instead, we can select two languages representing extremes in linguistic distance from American English. We chose African-accented English as the closest and Mandarin Chinese as one of the most linguistically distant. This selection is based on linguistic distance studies~\cite{chiswick2005linguistic}, making the evaluation of other languages like German or Spanish redundant, as they will not be closer than African-accented English and are less distant than Mandarin Chinese as shown in Figure \ref{fig:languages}.

\label{sec:format}

\noindent
\textbf{Downstream Task:}
As a downstream task, we use phoneme recognition for an American English dataset labelled with phoneme classes. Similarly to the APC model choice, our method can be easily extended to other downstream tasks such as speech recognition.

%\section{Methodology}
\section{Experimental Setup}
\subsection{Datasets}
To simulate low-resource scenarios we use the \libtrainclean{} partition from the Librispeech dataset~\cite{7178964} for pre-training. Librispeech is an American English reading speech database widely used in the literature for research benchmarks. A fixed random subset of 25-hrs is extracted from \libtrainclean{} for the baseline model. 
The phoneme recognition downstream task is performed by finetuning the APC model on fixed 10-hrs of speech extracted from Librispeech \libtraincleanother{}. It should be noted that we make sure that no data is overlapping between the finetuning and the pre-training datasets. 
To test the performance we use the Librispeech standard \libtestclean{} dataset. For both partitions \libtraincleanother{} and \libtestclean{} we take the phoneme labels from~\cite{Liu2019MockingjayUS}.

For the cross-lingual experiments, we extract African-accented English from the Zindi AfriSpeech competition\footnote{https://zindi.africa/competitions/intron-afrispeech-200-automatic-speech-recognition-challenge} and Mandarin Chinese from the Common Voice Chinese speech dataset\footnote{https://commonvoice.mozilla.org/en/datasets}. For data augmentation, we extract background noise from the MUSAN database~\cite{musan2015}.

\subsection{APC Architecture}
The architecture of the APC model is a Long Short-term Memory (LSTM) structure with skip connections \cite{HochSchm97}. The LSTM network consists of 3 layers. Each layer size is 512. The input to the APC model is a sequence of mel-spectrogram calculated with 80 mel bands while the output is 512-size feature representations. The training criteria of the APC model is the 3-step prediction of the future frames. Mean square error (MSE) loss is computed between prediction and ground truth frames. The APC model is pre-trained for 100 epochs with a learning rate of 0.0001 and batch size of 32. Adam optimizer is used for neural network optimization.

In the downstream phoneme recognition task, feed-forward linear layers are attached to the last layer of the APC model. The input dimension of the linear layer is 512 while the output dimension is the total number of phonemes. Log softmax is used to convert the output of the final layer to probabilities. Finally, the cross-entropy loss is computed between probabilities and ground truth phoneme classes. The remaining hyper-parameters are the same as~\cite{Chung2019AnUA}.

\subsection{Experiments}

\begin{table}[!t]
\caption{\label{tab:table1}Augmentation strategy on pre-training}
\centering
\begin{tabular}{lll} \\\hline
\textbf{Pre-training data}       & \textbf{No. of Hrs} & \textbf{Aug. ratio} \\\hline
Clean Data                       & 25,50,75,100        & N/A              \\
Clean25+[African]                & 50, 75, 100         & 1, 2, 3          \\
Clean25+[Chinese]                & 50, 75, 100         & 1, 2, 3          \\
Clean25+[Noise-aug]              & 50, 75, 100         & 1, 2, 3          \\
Clean25+[Pitch-aug]              & 50, 75, 100         & 1, 2, 3          \\
Clean25+[Noise/Pitch]            & 50, 75, 100, 175,   & 1, 2, 3, 6,       \\
                                 & 325,425,525     & 12,16,20       \\
\end{tabular}
\end{table}

Using the APC model, the following experiments are conducted with the respective datasets.
 
\noindent
\textbf{Clean Data}: The clean data experiment consists of mixing the baseline Librispeech 25-hrs with more data extracted from the same partition \libtrainclean{}. This experiment represents the oracle experiment, telling us how the model would behave if these data were available.

For the other scenarios, we mixed baseline Librispeech 25-hrs, referred to as Clean25, with augmented data as pre-training augmented data, specifically:

\noindent
\textbf{Clean25 + [African accented English Data]}: To Clean25, we added African-accented English speech data.

\noindent
\textbf{Clean25 + [Chinese speech Data]}: To Clean25, we added real Chinese speech data.

\noindent
\textbf{Clean25 + [Pitch-aug]}: We first applied synthetic pitch-modification effects to Clean25 and then combined Clean25 and pitch-augmented Clean25.

\noindent
\textbf{Clean25 + [Noise-aug]}: We added noise augmentations to Clean25 and then combined to Clean25 and the noise-augmented Clean25 in pre-training data.

\noindent
\textbf{Clean25 + [Noise-aug, Pitch-aug]}: We added pitch-modification and noise-augmentation effects to Clean25 and combined them in the pre-training data. We increased the pre-trained augmented data with more augmentation factors.

In all these experiments we kept the fine-tuning and evaluation data fixed. We only experimented and applied different augmentation strategies to the pre-training data.
Noise augmentation has been done by mixing training data with additive noise samples in the time domain with 5dB, 10dB and 15dB signal-to-noise (SNR) ratios. Pitch shifting is applied to the pre-training data so that the model learns pitch-invariant representations.
These experiments are summarised in Table~\ref{tab:table1}.

\begin{figure}[htb]
	
	\begin{minipage}[b]{1.0\linewidth}
		\centering
		\centerline{\includegraphics[width=0.95\linewidth]{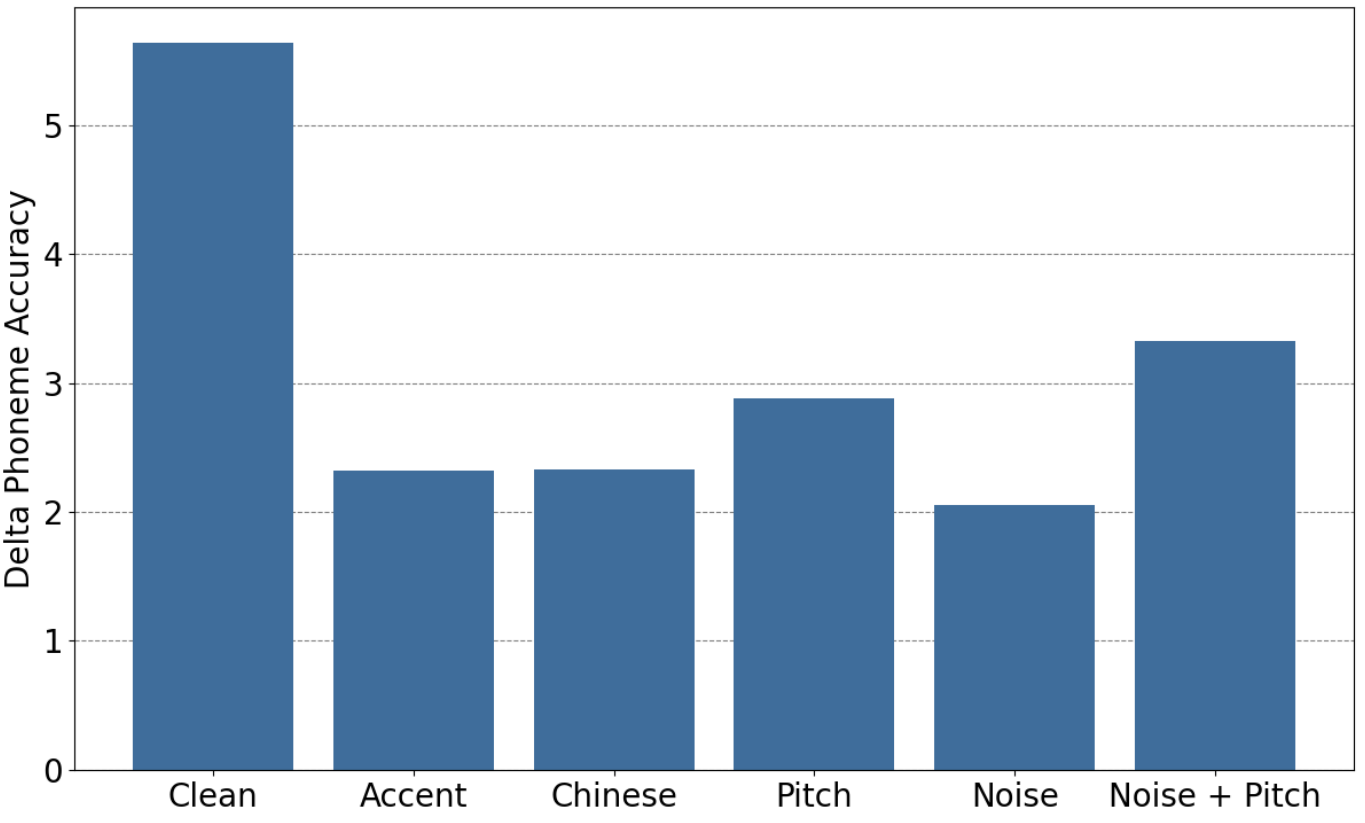}}
		%  \vspace{2.0cm}
	\end{minipage}
	\caption{Performance of 100-hrs augmented pre-training data (25-clean-hrs + 75-aug-hrs) compared to the baseline 25-clean-hrs (baseline acc. 51.5\%)}
	\label{fig:barplot}
	%
%\end{figure}
\end{figure}

%\begin{figure*}[hbt!]
\begin{figure*}[htb]
	
	\begin{minipage}[b]{1.0\linewidth}
		\centering
		\centerline{\includegraphics[width=0.85\linewidth]{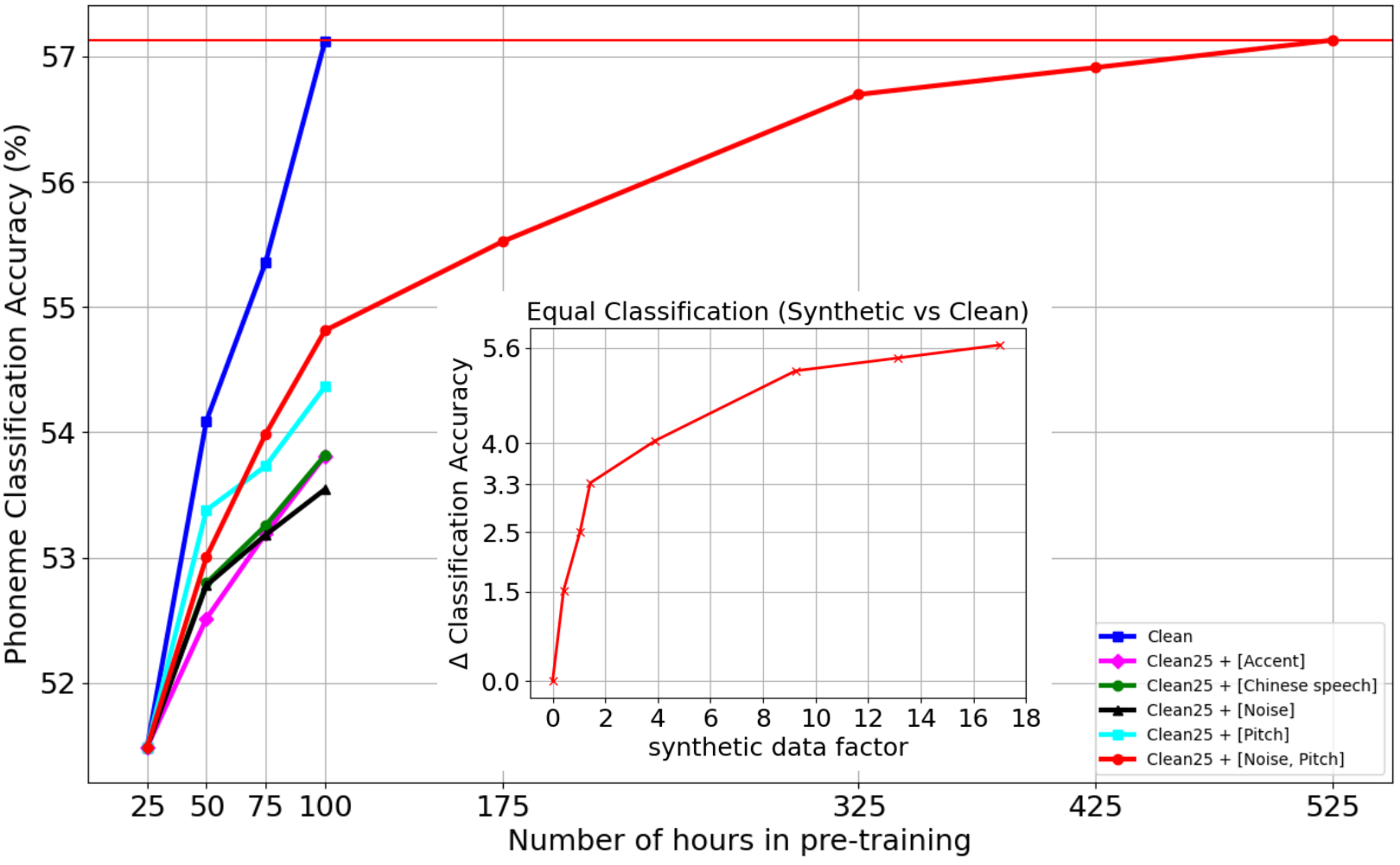}}
		%  \vspace{2.0cm}
	\end{minipage}
	\caption{Performance of augmented vs clean data in pre-training}
	\label{fig:results}
	%
%\end{figure}
\end{figure*}

\section{Results}
\subsection{Comparison of Augmentation Strategies}
First, we analyse the performance of all the augmentation strategies (artificial vs cross-lingual) by fixing to 75 hours of augmentation for each technique.
Figure~\ref{fig:barplot} shows the performance gain on top of the 25-hrs baseline accuracy (51.5\%) for each augmentation strategy (accent-aug, language-aug, noise-aug, pitch-aug, mix [pitch, noise]-aug) on Librispeech \libtestclean{} dataset. Figure~\ref{fig:barplot} shows that adding African-accented English and Mandarin Chinese speech data show a small improvement in the phoneme classification. Pitch augmentation performed better than augmenting with accented or other language data. The noise-augmentation also improves phoneme accuracy but less than the other strategies tested. However, Figure~\ref{fig:barplot} highlights that combining synthetic pitch-modification and noise augmentation performs better than the cross-lingual pre-training or individual artificial data augmentation.

\subsection{Augmentation Size Analysis}
A second analysis is conducted to understand the influence of the number of augmentation hours. In Figure~\ref{fig:results} we plot the phoneme classification (\%) against the total number of hours in pre-training data for each augmentation strategy. Instead of fixing the amount of augmentation (both artificial and cross-lingual) we progressively add 25, 50, and 75 augmented hours to understand how the size influences the pre-training phase.
Figure~\ref{fig:results} shows that adding audio augmentations consistently improves the performance for any augmentation size. This performance improvement is visible from the steep slopes of all augmentation approaches in Figure~\ref{fig:results}. The highest slope is obtained with the oracle model i.e., adding original data of the target low-resource language. Regarding the augmentation strategies, we observe that pitch modification works better than cross-lingual pre-training and that combining pitch and noise is the best augmentation strategy. These results show that artificial data augmentation is a useful technique for low-resource scenarios. In addition, we highlight that contrasting artificial data augmentation with cross-lingual pre-training is only made to contextualize our results. In practice, given these results, we suggest that artificial data augmentation could be also used together with cross-lingual pre-training. 

We extended our experiments by adding 150-hrs, 300-hrs, 400-hrs and 500-hrs [noise, pitch] augmentations to the pre-training data and performed the pre-training process as shown in Figure~\ref{fig:results}. Using 525-hrs mix of augmentation in pre-training data matches the performance of the oracle model, i.e. more data from Librispeech \libtrainclean{} as shown in red and blue plots in figure \ref{fig:results}.
From our results in Figure~\ref{fig:results}, we observed that adding augmentation to the baseline 25-hrs pre-training data improves the performance on downstream phoneme classification. To better understand how much augmentation data in pre-training improves the phoneme accuracy, the inset in Figure~\ref{fig:results} shows a graph for achieving a certain level of accuracy using our augmentation approach. The y-axis shows the delta phoneme accuracy (\%) between the standard Librispeech \libtrainclean{} and the baseline 25-hrs, while the x-axis shows the additional augmentation multiplier data used in pre-training data. With 75-hrs (3 times more data) added to the baseline (clean25), we observed delta phoneme classification improvement of 5.6\%. To achieve that level of phoneme accuracy, we need to add 17-times more augmentation data to the pre-training data. We also noted that adding 3-times synthetic augmentation achieved a 3.3\% improvement which is 60\% of the improvement compared to pre-training with the same amount of Librispeech data as shown in subplot Figure~\ref{fig:results}. For the same accuracy, we require 1.6-times more augmented data than using more pre-training data from Librispeech -- an option that might not be available in resource-constrained scenarios. We also noticed that the improvement due to augmentation is not consistent, the tail of graph becomes flat with more augmentation. This shows the limitations of augmentation strategy in pre-training stage.

\section{Conclusions}
We examined SSRL augmentation-based pre-training strategies where the amount of pre-training data is limited finding artificial data augmentation out-performed cross-lingual pre-training. We scaled the augmentation for the best strategy achieving equal performance to the model pre-trained with target speech data with 17-times augmentation.  
This paper highlights the potential of artificial data augmentation as an effective technique for resource-constrained SSRL pre-training, showing better results than equivalent transferring knowledge from alternative languages or accents to the target. The two techniques should not be seen as mutually exclusive as both techniques are beneficial. Our results indicate that on top of cross-lingual pre-training, researchers should also consider adding artificial data (up to 2-times) augmentation to further improve the results for low-resource languages. In future, we will consider adding synthetic speech generated from text-to-speech (TTS) systems in the SSRL model pre-training process and extend our work to more recent open-source pre-trained models.

\section{Acknowledgements}
This paper emanated from research funded by Science Foundation Ireland to the SFI Centre for Research Training in Machine Learning (18/CRT/6183) and Insight Centre for Data Analytics (12/RC/2289\_P2). For the purpose of Open Access, the author has applied a CC BY public copyright licence to any Author Accepted Manuscript version arising from this submission.

\bibliographystyle{IEEEtran}
\bibliography{mybib}

\begin{thebibliography}{10}
\providecommand{\url}[1]{#1}
\csname url@samestyle\endcsname
\providecommand{\newblock}{\relax}
\providecommand{\bibinfo}[2]{#2}
\providecommand{\BIBentrySTDinterwordspacing}{\spaceskip=0pt\relax}
\providecommand{\BIBentryALTinterwordstretchfactor}{4}
\providecommand{\BIBentryALTinterwordspacing}{\spaceskip=\fontdimen2\font plus
\BIBentryALTinterwordstretchfactor\fontdimen3\font minus \fontdimen4\font\relax}
\providecommand{\BIBforeignlanguage}[2]{{%
\expandafter\ifx\csname l@#1\endcsname\relax
\typeout{** WARNING: IEEEtran.bst: No hyphenation pattern has been}%
\typeout{** loaded for the language `#1'. Using the pattern for}%
\typeout{** the default language instead.}%
\else
\language=\csname l@#1\endcsname
\fi
#2}}
\providecommand{\BIBdecl}{\relax}
\BIBdecl

\bibitem{hammarstrom2005review}
H.~Hammarstr{\"o}m, ``Review of raymond j. gordon, jr.(ed.) 2005 ethnologue: Languages of the world, sil international,'' Citeseer, Tech. Rep., 2005.

\bibitem{moseley2010atlas}
C.~Moseley, \emph{Atlas of the World's Languages in Danger}.\hskip 1em plus 0.5em minus 0.4em\relax Unesco, 2010.

\bibitem{ting2010language}
D.~S. Ting and Y.-Y. Puan, \emph{Language attitudes of Hokkien speakers towards Hokkien and Mandarin}.\hskip 1em plus 0.5em minus 0.4em\relax Universiti Malaysia Sarawak, 2010.

\bibitem{rahman1995pashto}
T.~Rahman, ``The pashto language and identity-formation in pakistan,'' \emph{Contemporary South Asia}, vol.~4, no.~2, pp. 151--170, 1995.

\bibitem{Chung2019AnUA}
Y.-A. Chung, W.-N. Hsu, H.~Tang, and J.~R. Glass, ``An unsupervised autoregressive model for speech representation learning,'' in \emph{INTERSPEECH}, 2019.

\bibitem{oord2018representation}
A.~van~den Oord, Y.~Li, and O.~Vinyals, ``Representation learning with contrastive predictive coding,'' 2018.

\bibitem{10.1109/TASLP.2021.3122291}
\BIBentryALTinterwordspacing
W.-N. Hsu, B.~Bolte, Y.-H.~H. Tsai, K.~Lakhotia, R.~Salakhutdinov, and A.~Mohamed, ``Hubert: Self-supervised speech representation learning by masked prediction of hidden units,'' \emph{IEEE/ACM Trans. Audio, Speech and Lang. Proc.}, vol.~29, p. 3451–3460, oct 2021. [Online]. Available: \url{https://doi.org/10.1109/TASLP.2021.3122291}
\BIBentrySTDinterwordspacing

\bibitem{Yang2021SUPERBSP}
S.~wen Yang, P.-H. Chi, Y.-S. Chuang, C.-I. Lai, K.~Lakhotia, Y.~Y. Lin, A.~T. Liu, J.~Shi, X.~Chang, G.-T. Lin, T.~hsien Huang, W.-C. Tseng, K.~tik Lee, D.-R. Liu, Z.~Huang, S.~Dong, S.-W. Li, S.~Watanabe, A.~rahman Mohamed, and H.~yi~Lee, ``Superb: Speech processing universal performance benchmark,'' in \emph{Interspeech}, 2021.

\bibitem{babu2021xls}
A.~Babu, C.~Wang, A.~Tjandra, K.~Lakhotia, Q.~Xu, N.~Goyal, K.~Singh, P.~von Platen, Y.~Saraf, J.~Pino \emph{et~al.}, ``Xls-r: Self-supervised cross-lingual speech representation learning at scale,'' \emph{arXiv preprint arXiv:2111.09296}, 2021.

\bibitem{sellam2023squid}
T.~Sellam, A.~Bapna, J.~Camp, D.~Mackinnon, A.~P. Parikh, and J.~Riesa, ``Squid: Measuring speech naturalness in many languages,'' in \emph{IEEE International Conference on Acoustics, Speech and Signal Processing (ICASSP)}.\hskip 1em plus 0.5em minus 0.4em\relax IEEE, 2023, pp. 1--5.

\bibitem{baevski2020wav2vec}
A.~Baevski, Y.~Zhou, A.~Mohamed, and M.~Auli, ``wav2vec 2.0: A framework for self-supervised learning of speech representations,'' \emph{Advances in neural information processing systems}, vol.~33, pp. 12\,449--12\,460, 2020.

\bibitem{ko2015audio}
T.~Ko, V.~Peddinti, D.~Povey, and S.~Khudanpur, ``Audio augmentation for speech recognition,'' in \emph{Sixteenth annual conference of the international speech communication association}, 2015.

\bibitem{Park2019SpecAugmentAS}
D.~S. Park, W.~Chan, Y.~Zhang, C.-C. Chiu, B.~Zoph, E.~D. Cubuk, and Q.~V. Le, ``Specaugment: A simple data augmentation method for automatic speech recognition,'' in \emph{Interspeech}, 2019.

\bibitem{ko2017study}
T.~Ko, V.~Peddinti, D.~Povey, M.~L. Seltzer, and S.~Khudanpur, ``A study on data augmentation of reverberant speech for robust speech recognition,'' in \emph{2017 IEEE International Conference on Acoustics, Speech and Signal Processing (ICASSP)}.\hskip 1em plus 0.5em minus 0.4em\relax IEEE, 2017, pp. 5220--5224.

\bibitem{kharitonov2021data}
E.~Kharitonov, M.~Rivi{\`e}re, G.~Synnaeve, L.~Wolf, P.-E. Mazar{\'e}, M.~Douze, and E.~Dupoux, ``Data augmenting contrastive learning of speech representations in the time domain,'' in \emph{2021 IEEE Spoken Language Technology Workshop (SLT)}.\hskip 1em plus 0.5em minus 0.4em\relax IEEE, 2021, pp. 215--222.

\bibitem{zhu2022noise}
Q.-S. Zhu, J.~Zhang, Z.-Q. Zhang, M.-H. Wu, X.~Fang, and L.-R. Dai, ``A noise-robust self-supervised pre-training model based speech representation learning for automatic speech recognition,'' in \emph{IEEE International Conference on Acoustics, Speech and Signal Processing (ICASSP)}.\hskip 1em plus 0.5em minus 0.4em\relax IEEE, 2022, pp. 3174--3178.

\bibitem{wang2022improving}
H.~Wang, Y.~Qian, X.~Wang, Y.~Wang, C.~Wang, S.~Liu, T.~Yoshioka, J.~Li, and D.~Wang, ``Improving noise robustness of contrastive speech representation learning with speech reconstruction,'' in \emph{IEEE International Conference on Acoustics, Speech and Signal Processing (ICASSP)}.\hskip 1em plus 0.5em minus 0.4em\relax IEEE, 2022, pp. 6062--6066.

\bibitem{huang2022spiral}
W.~Huang, Z.~Zhang, Y.~T. Yeung, X.~Jiang, and Q.~Liu, ``Spiral: Self-supervised perturbation-invariant representation learning for speech pre-training,'' \emph{arXiv preprint arXiv:2201.10207}, 2022.

\bibitem{baevski2021unsupervised}
A.~Baevski, W.-N. Hsu, A.~Conneau, and M.~Auli, ``Unsupervised speech recognition,'' \emph{Advances in Neural Information Processing Systems}, vol.~34, pp. 27\,826--27\,839, 2021.

\bibitem{ullah2023improving}
A.~Ullah, A.~Ragano, and A.~Hines, ``Improving phoneme recognition with augmented autoregressive predictive coding,'' in \emph{2023 34th Irish Signals and Systems Conference (ISSC)}.\hskip 1em plus 0.5em minus 0.4em\relax IEEE, 2023, pp. 1--6.

\bibitem{Sriram2022Wav2VecAugIS}
A.~Sriram, M.~Auli, and A.~Baevski, ``Wav2vec-aug: Improved self-supervised training with limited data,'' in \emph{Interspeech}, 2022.

\bibitem{Chung2019GenerativePF}
Y.-A. Chung and J.~R. Glass, ``Generative pre-training for speech with autoregressive predictive coding,'' \emph{IEEE International Conference on Acoustics, Speech and Signal Processing (ICASSP)}, pp. 3497--3501, 2019.

\bibitem{Chung2020VectorQuantizedAP}
Y.-A. Chung, H.~Tang, and J.~R. Glass, ``Vector-quantized autoregressive predictive coding,'' \emph{ArXiv}, vol. abs/2005.08392, 2020.

\bibitem{chung-glass-2020-improved}
\BIBentryALTinterwordspacing
Y.-A. Chung and J.~Glass, ``Improved speech representations with multi-target autoregressive predictive coding,'' in \emph{Proceedings of the 58th Annual Meeting of the Association for Computational Linguistics}.\hskip 1em plus 0.5em minus 0.4em\relax Online: Association for Computational Linguistics, Jul. 2020, pp. 2353--2358. [Online]. Available: \url{https://aclanthology.org/2020.acl-main.213}
\BIBentrySTDinterwordspacing

\bibitem{DBLP:journals/jstsp/YangYCGT22}
\BIBentryALTinterwordspacing
G.~Yang, S.~Yeh, Y.~Chung, J.~R. Glass, and H.~Tang, ``Autoregressive predictive coding: {A} comprehensive study,'' \emph{{IEEE} J. Sel. Top. Signal Process.}, vol.~16, no.~6, pp. 1380--1390, 2022. [Online]. Available: \url{https://doi.org/10.1109/JSTSP.2022.3203608}
\BIBentrySTDinterwordspacing

\bibitem{hsu2021hubert}
W.-N. Hsu, B.~Bolte, Y.-H.~H. Tsai, K.~Lakhotia, R.~Salakhutdinov, and A.~Mohamed, ``Hubert: Self-supervised speech representation learning by masked prediction of hidden units,'' \emph{IEEE/ACM Transactions on Audio, Speech, and Language Processing}, vol.~29, pp. 3451--3460, 2021.

\bibitem{chiswick2005linguistic}
B.~R. Chiswick and P.~W. Miller, ``Linguistic distance: A quantitative measure of the distance between english and other languages,'' \emph{Journal of multilingual and multicultural development}, vol.~26, no.~1, pp. 1--11, 2005.

\bibitem{7178964}
V.~Panayotov, G.~Chen, D.~Povey, and S.~Khudanpur, ``Librispeech: An asr corpus based on public domain audio books,'' in \emph{IEEE International Conference on Acoustics, Speech and Signal Processing (ICASSP)}, 2015, pp. 5206--5210.

\bibitem{Liu2019MockingjayUS}
A.~T. Liu, S.~wen Yang, P.-H. Chi, P.-C. Hsu, and H.~yi~Lee, ``Mockingjay: Unsupervised speech representation learning with deep bidirectional transformer encoders,'' \emph{IEEE International Conference on Acoustics, Speech and Signal Processing (ICASSP)}, pp. 6419--6423, 2019.

\bibitem{musan2015}
D.~Snyder, G.~Chen, and D.~Povey, ``{MUSAN}: {A} {M}usic, {S}peech, and {N}oise {C}orpus,'' 2015, arXiv:1510.08484v1.

\bibitem{HochSchm97}
S.~Hochreiter and J.~Schmidhuber, ``Long short-term memory,'' \emph{Neural Computation}, vol.~9, no.~8, pp. 1735--1780, 1997.

\end{thebibliography}

\end{document}